\documentclass[final,5p,times,twocolumn,authoryear]{elsarticle}

\usepackage{amssymb}
\usepackage{graphicx}
\usepackage{subcaption}
\usepackage{natbib}
\usepackage{caption}
\usepackage{color}
\usepackage{xcolor}

\usepackage{hyperref}
\hypersetup{colorlinks=true,
            linkcolor=blue,
            anchorcolor=blue,
            citecolor=blue}

\definecolor{dark_red}{HTML}{8B0000}
\newcommand {\rev}[1]{#1}

\journal{Visual Informatics}
\begin{document}

\begin{frontmatter}

\title{Visual Analysis of LLM-based Entity Resolution from Scientific Papers}

\cortext[cor]{Corresponding author}
\fntext[equalcon]{These authors contributed equally to this work.}

\author[Software]{Siyu Wu\fnref{equalcon}}

\author[ComputerScience]{Yi Yang\fnref{equalcon}}

\author[ComputerScience]{Weize Wu}

\author[Software]{Ruiming Li}
\author[Software]{Yuyang Zhang}

\author[USTB]{Ge Wang}

\author[Software]{Huobin Tan}

\author[Software]{Zipeng Liu\corref{cor}}
\ead{zipeng@buaa.edu.cn}

\author[ComputerScience]{Lei Shi\corref{cor}}
\ead{shijim@gmail.com}

\affiliation[Software]{organization={School of Software, Beihang University},
            city={Beijing},
            country={China}}

\affiliation[ComputerScience]{organization={School of Computer Science and Engineering, Beihang University},
            city={Beijing},
            country={China}}

\affiliation[USTB]{organization={School of Materials Science and Engineering, University of Science and Technology Beijing},
            city={Beijing},
            country={China}}

\begin{abstract}
This paper focuses on the visual analytics support for extracting domain-specific entity from extensive scientific literature, a task with inherent limitations using traditional named entity resolution methods. With the advent of large language models (LLMs) such as GPT-4, significant improvements over conventional machine learning approaches have been achieved due to LLM's capability on entity resolution integrate abilities such as understanding multiple types of text.
This research introduces a new visual analysis pipeline that integrates these advanced LLMs with versatile visualization and interaction designs to support batch entity resolution.
Specifically, we focus on a specific material science field of Metal-Organic Frameworks (MOFs) and a large data collection namely CSD-MOFs. Through collaboration with domain experts in material science, we obtain well-labeled synthesis paragraphs.
We propose human-in-the-loop refinement over the entity resolution process using visual analytics techniques, which allows domain experts to interactively integrate insights into LLM intelligence, including error analysis and interpretation of the retrieval-augmented generation (RAG) algorithm. Our evaluation through the case study of example selection for RAG demonstrates that this human-machine collaborative approach improved single-document entity resolution accuracy by approximately 30\%.
\end{abstract}


\end{frontmatter}

\section{Introduction}
\label{sec:Introduction}

Entity resolution from scientific papers plays a crucial role in advancing our understanding and organization of scientific knowledge \citep{li2020survey}. Unlike plain text or unstructured data, entity resolution from scientific papers often preserves rich structural relationship that enables more sophisticated downstream applications \rev{such as identifying research trends \citep{marrone_application_2020} and building biomedical knowledge base \citep{zheng_entity_2015}}. Such structured information is particularly valuable for constructing domain-specific knowledge databases. Moreover, the relationships between extracted entities can be integrated into higher-level semantic representations, \rev{such as detailed synthesis recipes (e.g., Cu and Zn in metals, H$_2$O and C$_2$H$_5$OH in solvents) and experimental operations (e.g., heating and vibration) for materials science} \citep{lyu2020digital}. However, the precise entity resolution from scientific papers presents significant challenges due to the inherent variability in literature structures and the diversity of writing styles across different authors.

Traditionally, entity resolution methods have depended on expert systems and conventional machine learning technologies \citep{li2020survey}. For instance, the DigiMOF database \citep{glasby_digimof_2023} highlights the challenges in manual data extraction from unstructured scientific literature. Despite advancements like \rev{ChemDataExtractor \citep{ChemDataExtractor-v2}}, the diverse sentence structures and terminologies in specific domains, such as \rev{Metal-Organic Frameworks (MOFs)} synthesis, lead to issues in data consistency and accuracy. This underscores the difficulty of relying solely on rule-based systems for complex scientific data.

\rev{In recent years, large models such as GPT-4 \citep{achiam2023gpt} were known to exhibit straightforward question-and-answer interface and high accuracy.} They have been recognized as revolutionary methods for literature analysis and are increasingly being applied in scientific research \citep{bubeck2023sparks}. Large language models offer significant advantages over traditional approaches in scientific entity resolution. They demonstrate superior performance across diverse text formats and can be easily adapted to new domains through techniques like prompt engineering and few-shot learning \citep{ai4science2023impact}.
A new challenge is observed in generative AI systems like ChatMOF \citep{kang2023chatmof}, which aims to predict and generate new high-performance MOFs. While it leverages large language models (LLMs) for reasoning and data integration, the inherent complexities of material science data often require supplementary tools for precise property predictions and error-prone iterative adjustments, particularly in structure generation tasks.

Despite the advantages of LLMs in scientific entity resolution, recent studies have highlighted persistent challenges in their application to scientific literature analysis. These include difficulties in accurate numerical data extraction, inconsistencies in structured output generation like material synthesis parameters, and limitations in complex reasoning tasks \citep{shi_comparison_2024}. In addition, more intuitive visualization and the capability to analyze resolution errors interactively are crucial to improving system performance and understanding failure cases.

\begin{figure*}[t]
    \centering
    \includegraphics[width=1\linewidth]{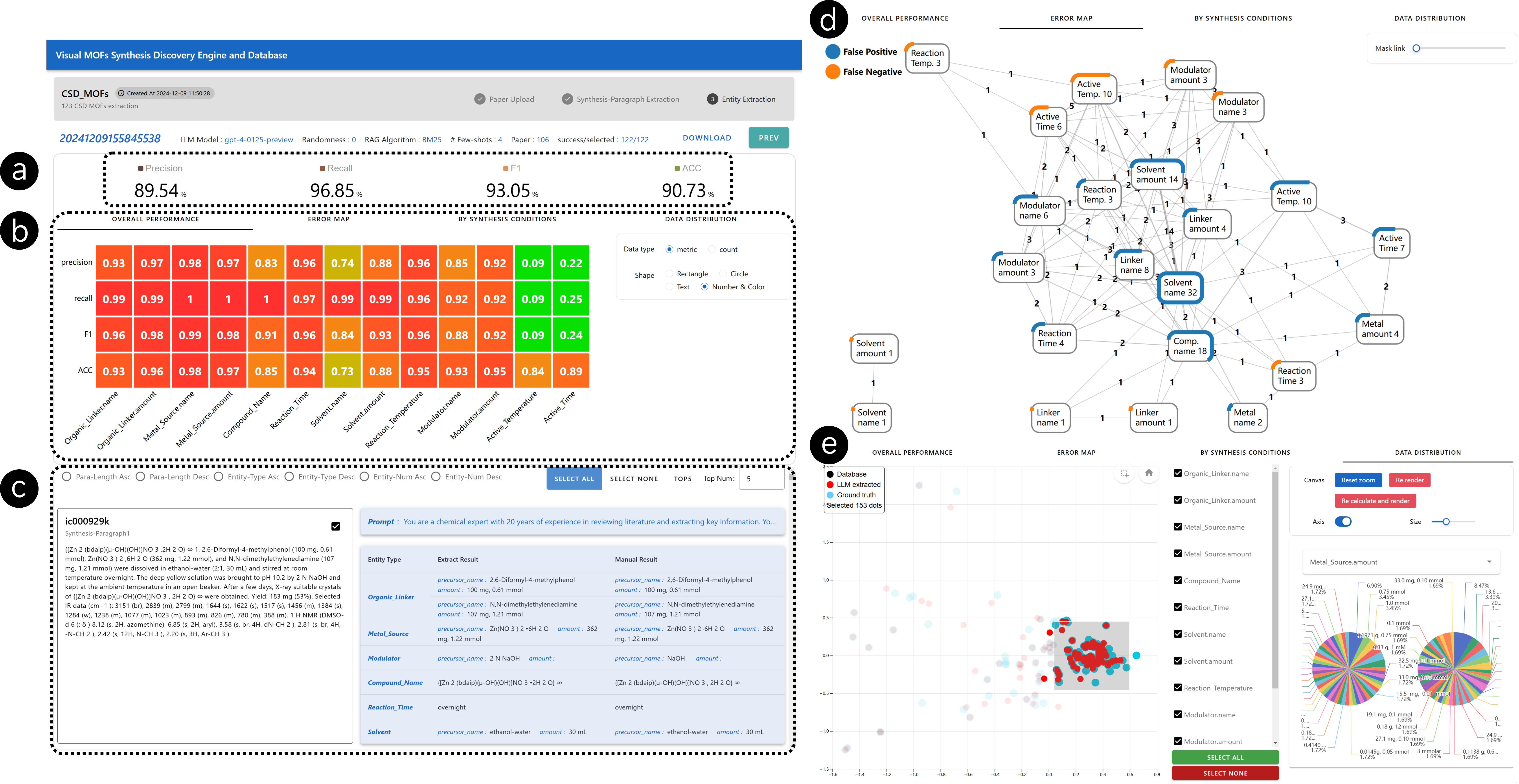}
    
    \caption{Overview of visualization interface.
    \rev{View (a) shows} 4 metrics including \textit{Precision, Recall, F1 and ACC} that characterize model performance.
    \rev{View (b)} is a heat map showing the 4 metrics on each parameter and user can switch the data type to ``count" to show TP, TN, FP, FN count on each parameter. 
    \rev{View (c)} is a filter panel that user can select paragraphs of interest. User can scroll down to view detailed information of all paragraphs.
    \rev{Users can switch to two additional views through the tab above View (b): View (d)} displays a network graph where each rounded rectangle represents an error type (e.g., false positive count on solvent name) and its colored border represents how many times this type of error has occurred, and \rev{View (e)} contains a scatter plot and two pie charts, representing the distribution of all synthesis paragraphs and the distribution of LLM extracted results and ground truth in selected paragraphs on specific parameter respectively.}
    \label{fig:main-view}
\end{figure*}

\begin{figure*}[t]
    \centering
    \includegraphics[width=0.9\linewidth]{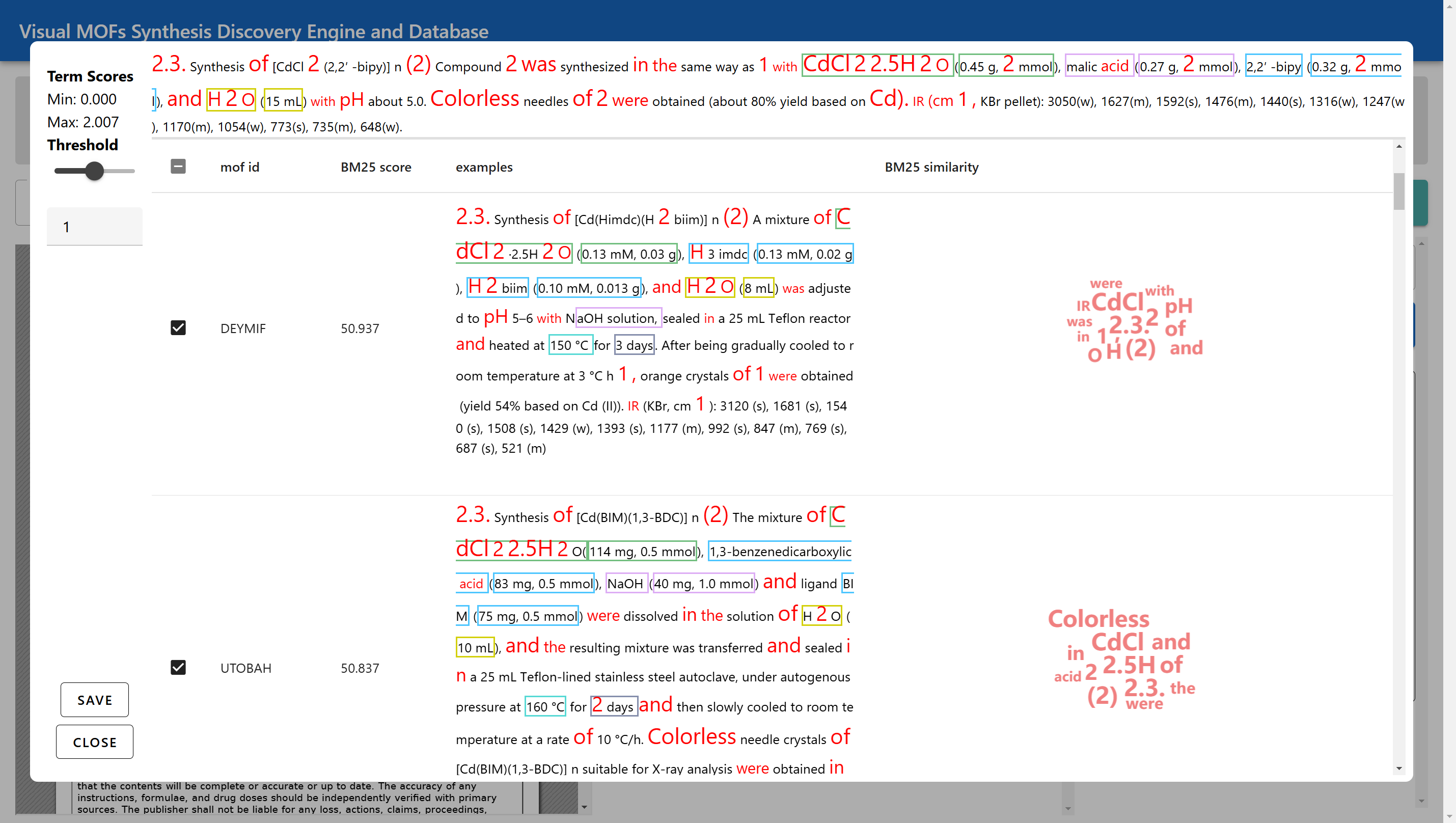}
    \caption{RAG algorithm configuration panel. The upper part of the panel shows the synthetic paragraph to be processed for entity resolution, while the lower panel presents several candidate paragraphs as examples. The red text within the paragraphs indicates matching text utilized by the BM25 algorithm. Users can apply paragraphs to RAG algorithm by selecting checkboxes located at the left of the candidate paragraphs.}
    \label{fig:rag}
\end{figure*}

Therefore, our research idea is to leverage the powerful capabilities of LLMs in comprehending diverse types of text and apply them to scientific entity resolution. Through collaboration with domain experts in materials science, we propose a visual analytics approach using Metal-Organic Frameworks (MOFs) as a case study and implement a series of visualization design, as shown in Fig.\ref{fig:main-view} and Fig.\ref{fig:rag}. Our approach integrates interactive visualization designs and interfaces to support three key aspects of LLM-based entity resolution: visualization of extraction results and performance metrics, error analysis of LLM entity resolution, and interpretation of the Retrieval-Augmented Generation (RAG) process in entity resolution. We evaluate the effectiveness of our approach through three case studies, demonstrating how our visual design can enhance the accuracy and interpretability of LLM-based entity resolution in scientific literature.

In comparison to existing methods, our work extends existing research in this field through the following approaches:

\begin{itemize}
\item \textbf{Technical Pipeline:} We have developed an integrated pipeline that utilizes LLMs for fully interactive, automated entity resolution over scientific literature. This pipeline enables users without any programming background to perform high-throughput entity resolution in their respective domains.

\item \textbf{Visualization and Interaction Design:} We have implemented a suit of visualization designs for the analysis of entity resolution results, transforming the human-machine dialogue interface of the large model into a visual analysis interface. This allows users to dynamically manipulate model parameters and view the visualized outcomes, thereby providing a clear, intuitive understanding of the entity resolution results.

\item \textbf{AI for Science Application:} We demonstrate the effectiveness of our framework within the material science domain through an application that showcases its ability to enable entity resolution across various document formats. We conducted an entity resolution on Metal-Organic Frameworks (MOFs) synthesis literature from CSD-MOFs dataset, comparing the resolution accuracy after cross-validation. The results achieved an overall accuracy of 90.73\% on well-labeled 122 synthesis paragraphs.
\end{itemize}

\section{Related Work}
\label{sec:related-works}

\rev{Our research builds upon prior work in LLM-based approaches for structured information extraction and visual analytics systems for enhancing machine learning (ML) methods through human-in-the-loop refinement.}

\rev{
\subsection{LLM-based Approaches for Structured Information Extraction}
}
\label{sec:rw:LLM-in-literature-anly}

Considerable research has been applied to using large language models (LLMs) across various scientific domains \citep{ai4science2023impact,bubeck2023sparks}, \rev{offering significant potential for extracting structured information from scientific texts} but facing challenges such as misalignment with sequence labeling, reduced accuracy, fictitious responses, and inconsistent outputs.

\cite{wang2023gpt} redefined entity labeling as a generative task by introducing identifiers to outputs, enhancing accuracy but increasing the need for extensive data processing.
\cite{dunn2022structured} demonstrated that LLMs, with adequate fine-tuning, can produce structured outputs facilitating structured extraction in literature analysis.
\cite{dagdelen2024structured} found that extensive fine-tuning enables LLMs to perform both entity and relationship extractions, further enhanced by human-in-the-loop feedback.
\cite{polak2024extracting} observed that in zero-shot scenarios, a meticulous combination of prompt engineering significantly enhances single-query accuracy, similar to the ``flow prompt" concept \citep{ridnik2024code}, yet dependent on complex querying workflows.
\cite{polak2023flexible} showed that zero-shot structured extraction, enhanced by human feedback or fine-tuning, enables effective extractions with minimal data. 
\cite{zheng2023chatgpt} reported that optimal prompt engineering allows LLMs to conduct effective zero-shot entity resolution and circumvent hallucinations, substituting traditional extensively trained models with adaptable LLMs. 

\rev{
In the materials science domain specifically, \cite{kang2023chatmof} created a question-answering agent using databases and LLMs for querying material science knowledge via single-query interactions, as an application of the agent method \citep{wang2024survey}.

Our work builds upon these domain-specific applications by developing an integrated visual analytics system that combines LLMs with interactive visualization to support batch entity resolution specifically for MOFs synthesis literature.
}

\rev{
\subsection{Visual Analytics for Interpreting the Results of ML Methods in NLP domain}
}
\label{sec:rw:data-visualization}

\rev{Visual analytics systems have emerged as powerful tools for interpreting and refining ML model outputs \citep{wang2023vis+}. Several recent systems have focused specifically on interpreting ML outputs through interactive visualization in natural language processing (NLP) domain.}

\cite{heimerl_embcomp_2022} introduced a robust visualization system that leverages various graphs to compare embedding and explore relationships within embedding spaces, marking a significant advancement in this area.
\rev{\cite{chang_2020_hotel_rev} employed visualization method to interpret the patterns of hotel reviews and responses analyzed through deep learning techniques.}

\rev{Particularly relevant to our work are visual analytics approaches that support human-in-the-loop refinement of ML outputs.}
\cite{reif_LinguisticLens_2023} presented a tool that expedites the analysis of LLM-generated text corpora, providing rapid insights into data structure and distribution.
\rev{\cite{Kahng2024} provided a convenient tool to compare the output results of different LLMs and help users intuitively understand the differences between one LLM and the baseline model.}
\cite{liu_jarvix_2023} designed a user-friendly tool that simplifies data summarizing with LLMs, requiring minimal user input and no coding, emphasizing the ease of generating accessible data insights.
\rev{\cite{zhao_lightva_2024} uses LLM agent to plan and execute data visualization tasks. This tool supports user interactive exploration to make LLM output closer to the goal of data exploration.}

\rev{Our work extends these approaches by specifically focusing on LLM-based entity resolution from scientific literature and integrating visualization techniques.}

\section{Background and Problem Definition}
\label{sec:bg-prob-def}

\subsection{LLM-based Entity Resolution from Scientific Papers}
\label{sec:bg:LLM-based-ER}

Scientific literature serves as a critical repository of domain-specific knowledge \citep{zheng_entity_2015}. However, the inherent diversity and structural complexity of these documents present challenges in extracting and structuring this knowledge. Entity resolution—the process of identifying, extracting, and linking semantically similar entities—is essential for constructing structured databases from unstructured text.

Traditional methods for entity resolution heavily rely on rule-based systems or machine learning techniques that demand significant domain expertise and manual effort \citep{Papadakis2020}. These approaches often involve crafting complex parsing rules or training models from scratch, both of which are computationally intensive and lack flexibility for adapting to new domains.

Large Language Models (LLMs), such as GPT-4, have transformed this field by demonstrating notable improvements in processing unstructured and diverse textual formats. Unlike traditional models, LLMs leverage extensive pre-trained knowledge, enabling them to perform entity resolution with minimal additional training \citep{dagdelen2024structured}. They can adapt to new domains through methods like zero-shot or few-shot learning, significantly reducing the reliance on domain-specific labeled data. For example, in material science, LLMs can extract entities such as material names, synthesis methods, and performance metrics directly from literature.

\rev{Despite these advancements of LLMs themselves, due to the diversity of writing styles, entity names (e.g., the names of various chemical materials used in MOFs synthesis) and the lack of high-quality well labeled training datasets, the deployment of LLM-based entity resolution workflows in real-world scientific domains remains limited.}

In this study, we design and implement an LLM-based entity resolution pipeline. Using Metal-Organic Frameworks (MOFs) for case studies, we demonstrate how LLMs can accurately extract and structure knowledge from scientific papers. The pipeline integrates advanced annotation techniques and utilizes LLMs to process large volumes of text, extract structured knowledge, and provide actionable insights for researchers.

\subsection{Visual Analysis Requirements}
\label{sec:bg:visual-based-eval}

\rev{
Although LLMs demonstrate significant potential for extracting structured knowledge from scientific literature, evaluating and refining their performance presents substantial challenges for domain experts.

Current evaluation approaches primarily rely on statistical metrics such as precision, recall, and F1 score, which quantify overall extraction accuracy but fail to provide actionable insights into specific error patterns or model limitations. In the context of materials science, domain practitioners often need to iteratively check and refine extraction results by analyzing error patterns and strategically selecting exemplary paragraphs to include in the LLM's context window for retrieval-augmented generation.

However, without specialized visualization tools, this process becomes unnecessarily cumbersome. This limits domain experts' ability to effectively leverage their specialized knowledge to improve extraction performance.
}

To address these challenges, we propose a visual-based evaluation framework for LLM-based entity resolution. This framework allows users to:

\begin{itemize}
    \item Observe the extraction results directly through intuitive visualizations that highlight the model’s performance and limitations.
    \item Interactively analyze errors and explore relationships among extracted entities.
    \item \rev{Visually assess the algorithms' limitations} by mapping its decision-making process onto interpretable graphs and charts.
\end{itemize}

For instance, in the context of MOF literature, we visualize entity resolution accuracy by cross-referencing extracted synthesis methods with known datasets. Interactive error analysis tools enable users to identify specific cases of under-performance, such as incorrect entity disambiguation or inconsistent resolution across similar documents. These tools also provide insights into correlations between input characteristics and model performance, facilitating opportunities for refinement.

By incorporating visual analysis tools into the LLM-based workflow, we bridge the gap between automated extraction and human interpretability. This approach enhances the usability of LLMs for domain practitioners and accelerates the adoption of these technologies in practical applications.

\subsection{Target User}
\label{sec:bg:target-user}

The proposed framework is tailored to serve three primary user groups:

\begin{itemize}
    \item \textbf{Domain Practitioners:} Researchers and professionals in specific fields, such as material science, who require structured knowledge extracted from extensive literature. For example, material scientists studying MOFs can utilize the framework to extract synthesis methods and performance metrics without needing extensive computational expertise. The interactive visualization tools empower these users to validate and refine extracted data effectively.
    \item \textbf{LLM Algorithm Developers:} Machine learning researchers and developers focused on enhancing LLM performance for specific tasks. The visual-based evaluation framework provides detailed insights into the model’s limitations and error patterns, enabling developers to optimize the models more efficiently.
    \item \textbf{NLP Researchers:} Specialists in natural language processing who explore the application of LLMs in scientific domains. The framework serves as a practical testbed for evaluating new techniques, such as prompt engineering and fine-tuning, in real-world contexts. Additionally, the visualization tools facilitate the exploration of model behavior and its adaptation to domain-specific tasks.
\end{itemize}

By addressing the needs of these user groups, the framework ensures broad applicability and fosters collaboration between domain practitioners and AI researchers. This integration of automated extraction, visualization, and user interaction represents a substantial advancement in the field of literature analysis.

\section{System Overview}
\label{sec:framework-overview}

\begin{figure*}[t]  
    \centering  
    \includegraphics[width=0.75\linewidth]{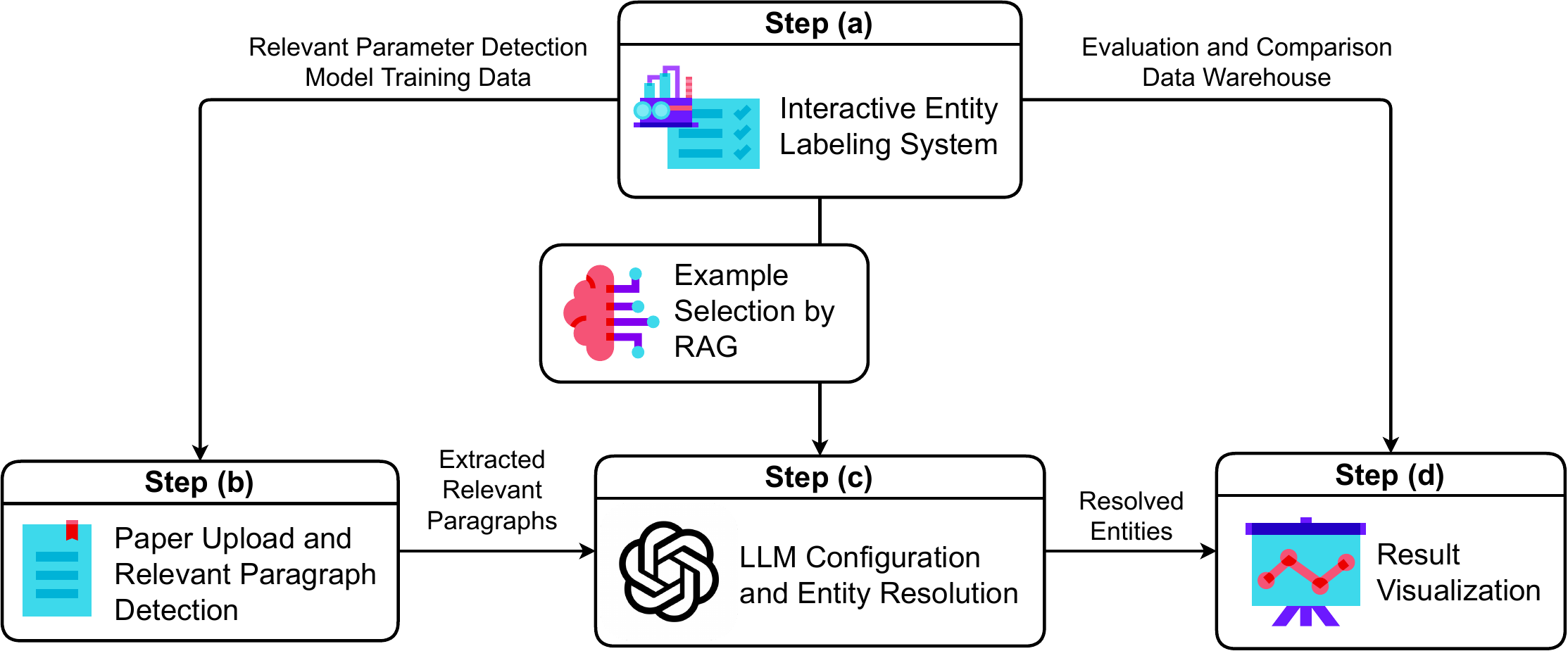}  
    \caption{Our system comprises four main components:
    (a) The labeling system construct high-accuracy data warehouse.
    (b) Users upload papers, which the framework then converts into paragraphs that encapsulate critical entity information, known as relevant paragraphs.
    (c) The interactive, user-friendly interface employs the LLM to derive structured entity resolution outcomes.
    (d) The framework processes and visualizes the structured data output.}
    \label{fig:overall_process} 
\end{figure*}

The framework for resolving entities in scientific papers integrates various innovative tools to provide a comprehensive, interactive visual analysis for literature entity resolution. This platform is characterized by its ease of use, batch resolution, comprehensive tool integration, few-shot LLM-based entity resolution, and visual analysis capabilities.
 
As shown in Fig.\ref{fig:overall_process}, the overall process of the platform involves several sequential phases including Interactive Entity Labeling System, Paper Upload and Relevant Paragraph Detection, LLM configuration and Entity Resolution, and finally, Results Visualization. 

\rev{
\subsection{Data Labeling and Entity Resolution Pipeline}
}
\label{sec:sys:pre-process}

The Entity Resolution workflow utilizes a labeled data warehouse throughout its three stages, supported by an interactive labeling system as shown in Step (a) of Fig.\ref{fig:overall_process}.

\rev{
In Step (a), we first collect data from CSD-MOFs and researchers label papers within a designated scientific domain after acquiring the PDF library. Administrators set up the entities to be labeled and delegate tasks. To ensure accuracy and consistency, annotators, experts in the specific field, are asked to label all the paragraphs related to MOFs synthesis in each paper, and their results are cross-validated. Then, data annotators are required to mark all entities in the paragraphs of MOFs synthesis process, including
\textit{organic linker name\&amount},
\textit{metal name\&amount},
\textit{solvent name\&amount},
\textit{modulator name\&amount},
\textit{compound name},
\textit{reaction time\&temperature},
and \textit{active time\&temperature}.
The result is JSON format, where the key is the entity name, and the value is the entity content. This process is iterative, refining labeling protocols based on feedback to optimize both efficiency and accuracy. Finally, we get 122 well-labeled synthesis paragraphs.

During the \textit{Paper Upload and Relevant Paragraph Detection} stage (Step b), our system receives PDF papers uploaded by users and then use an offline moderate machine learning model to classify whether a paragraph describes MOFs synthesis, which is convenient for us to perform the task of entity resolution.

In the \textit{LLM Configuration and Entity Resolution} stage (Step c), this data generate samples that include relevant paragraphs and entities for the LLM. The final stage, \textit{Result Visualization} (Step d), uses the labeled database for visual comparison with resolved entities.
}

\subsection{LLM Configuration}
\label{sec:sys:LLM-extract}

\begin{figure}[htbp]
    \centering
    \includegraphics[width=0.9\linewidth]{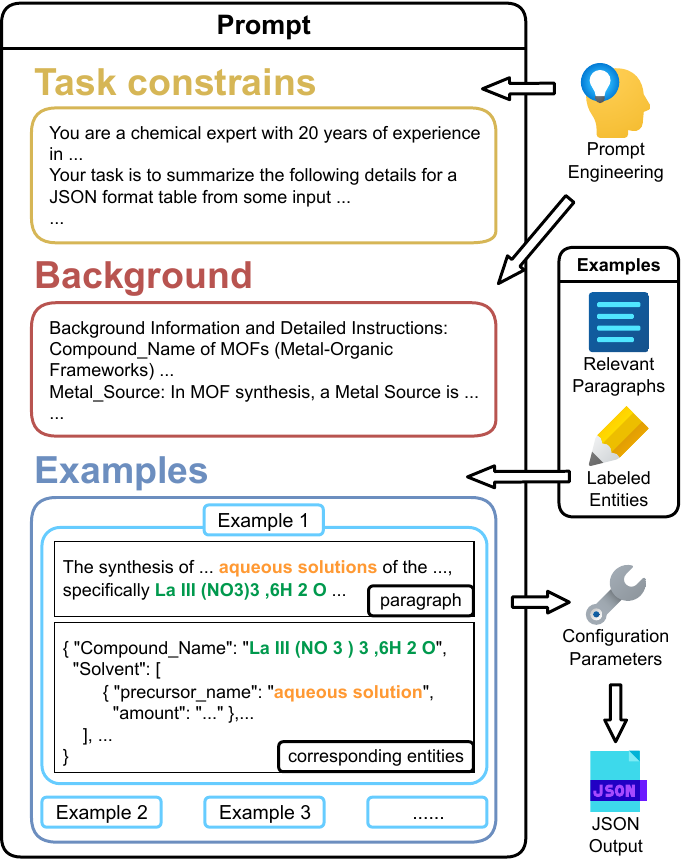}
    \caption{\rev{
        Prompt engineering of LLM-based entity resolution task in the MOFs example.
        It provides a way to interact with LLM, which can help build the prompt framework that users input to LLM.
        \textit{Paragraphs and labeled entities examples} are produced from our manually-labeled data, serving as a supplement to background knowledge and operation guide.
        \textit{Configuration parameters} of the model are adjustable to optimize the performance of LLM's output. For example, the user can set the temperature and the number of examples to check output results.
    }}
    \label{fig:prompt-template}
\end{figure}

The framework controls the output by manipulating prompts and model parameters. As shown in Fig.\ref{fig:prompt-template}, the prompts include prompt engineering, labeled entity resolution examples, and relevant paragraphs.

\textbf{Prompt Engineering} facilitates the content generation of LLMs, similar to programming in natural language, thus increasing accuracy and minimizing hallucinations \citep{white2023prompt, white2023chatgpt,bsharat2023principled}. After composing a prompt engineering strategy for entity resolution tasks, it proves universal across different scientific fields and LLMs, despite customized prompt engineering can enhance accuracy. Additionally, this approach incorporates structured commands in JSON format, ensuring the output remains well-organized for straightforward parsing and analysis by the platform. 
\rev{Our full prompt are available on \href{https://arxiv.org/abs/2408.04665}{arXiv} \citep{shi2025llm-mof}.}

\textbf{Labeled Entity Resolution Samples} are produced from labeled data in Step (a) of Fig.\ref{fig:overall_process}, with the platform presetting the generation quantity but allowing user customization. Each sample comprises a relevant paragraph and its corresponding entities, formatted in JSON within the prompt words.

\textbf{Relevant Paragraphs} were extracted from user-submitted papers in Step (b) of Fig.\ref{fig:overall_process}. \rev{Only those paragraphs containing essential information about MOFs synthesis are used, helping to reduce the length of input context to LLM.}
These paragraphs are first converted from PDF to HTML with the pdf2htmlEX tool \citep{wang2013online}, and then to plain text by removing formatting information. A specifically trained discrimination model then identifies key entity-containing paragraphs from these converted texts. This model, tailored for specific scientific domains and trained on data from Step (a) of Fig.\ref{fig:overall_process}, utilizes the labeled data for precise identification.
For instance, in the MOF domain, we refined a synthesis paragraph discrimination model using the bert-base-uncase model from Hugging Face \citep{DBLP:journals/corr/abs-1810-04805}, achieving an F1 score of 0.95, demonstrating exceptional accuracy.

\textbf{Configuration Parameters} of the model are adjustable to tailor the output behavior specific to the model used. For platforms employing OpenAI's GPT-3.5 and GPT-4.0 \citep{brown2020language, achiam2023gpt}, the temperature setting controls the creativity and randomness of responses. Configurations in LLMs from other developers vary due to unique training and deployment practices \citep{team2023gemini, touvron2023llama}, making them not as universally applicable as other configurations, thus are subject to ongoing optimization in our system.
\rev{The user can also decide how many examples to input to LLM. Then, we use retrieval-augmented generation (RAG) algorithm to select a user specified number of samples from our labeling system as the sample part of the input of LLM.}

\rev{
The testing data used to identify the optimized configuration is the MOFs synthesis paragraphs from CSD-MOFs dataset we labeled. We employed optimization techniques by systematically evaluating different configurations. We compared three RAG algorithms (BM25, BERT, and Sentence-BERT) and varied the number of few-shot examples to determine the best-performing setup. For the RAG algorithms, we selected BM25 as the optimal choice based on performance metrics (F1 and ACC). More technical details are available on \href{https://arxiv.org/abs/2408.04665}{arXiv} \citep{shi2025llm-mof}.
In our pipeline, we set the default temperature to 0 and the default sample size to 4.
}

\rev{The well labeled dataset, synthetic paragraph classification model, RAG algorithm, and the precise combination of model temperature and sample size ensure the high accuracy of LLM output results and reduce hallucinations.}
Users with a background in related science domains or machine learning can review and interactively adjust these settings to enhance resolution accuracy and further analyze and experiment with how different extraction parameters affect the results.

\subsection{LLM Performance Evaluation}
\label{sec:sys:LLM-eval}

After the LLM completes the entity resolution process, the platform displays and visually analyzes the results in the user interface, as depicted in Fig.\ref{fig:main-view}. The backend processes and organizes structured outputs for each resolved paper, matching each entity with its corresponding quantity and adjusting the display based on the entity's presence and count. If resolved papers are in the database, the system displays an accuracy matrix through heatmaps and a network diagram, which users can toggle to suit their needs. Additionally, the platform performs a dimensionality reduction analysis, visually highlighting the similarities and differences with other data in the database.

In a case study involving 106 fully labeled papers, the system compares LLM extracted result with existing ground truth using a measurement matrix that includes True Positive (TP), False Positive (FP), True Negative (TN), and False Negative (FN) metrics to calculate Accuracy, Precision, Recall, and F1 scores. Notably, if a labeled relevant paragraph lacks an entity, each non-resolution is counted as one TN. These metrics account for the non-existence of expected entities and incorrect resolutions, aiming to enhance model accuracy and mitigate the generation of unfounded results—a common issue with LLMs in scientific applications \citep{ai4science2023impact}.

\section{Visualization Design}
\label{sec:visualization}

\rev{
The visualization system was designed through close collaboration with domain experts in materials science to address their specific needs when working with LLM-based entity resolution from scientific literature. Through iterative design sessions and expert feedback, we developed an interface that helps domain practitioners visually understand LLM's performance in extracting synthesis parameters from MOF paragraphs from multiple perspectives.

The system enables experts to compare extraction performance across different synthesis parameters, facilitating the identification of specific shortcomings in LLM's extraction capabilities for particular entity types. This human-centered design approach ensures that the visualization effectively supports the analytical workflows of materials scientists while providing appropriate levels of detail for both overview analysis and in-depth investigation of extraction results.
}

\subsection{Example Selection in RAG}
\label{sec:vis:para-select}

The interface (Fig.\ref{fig:rag}) highlights terms and phrases identified by the model as relevant to synthesis, using color-coding such as red or yellow to indicate different terms. It also displays alternative examples with similarity scores calculated by BM25 algorithm to quantify relevance, allowing users to select or deselect examples to refine the model. A threshold adjustment slider enables users to fine-tune the criteria for significant term selection.

\subsection{Result Visualization}
\label{sec:vis:result-visual-panel}

The main view of the visualization interface, as shown in Fig.\ref{fig:main-view}, is composed of three parts.
\rev{From top to bottom they are} \textbf{overall metrics view} (Fig.\ref{fig:main-view} a), \textbf{detailed result view} (Fig.\ref{fig:main-view} b), and \textbf{extracted paragraphs view} (Fig.\ref{fig:main-view} c).
\rev{This interface structure deliberately organizes information density from abstract to concrete as users scan from top to bottom, enabling them to easily locate areas of interest.}

~\\\noindent \textbf{Overall Metrics View}\hspace{1em}

This view shows the model's extraction performance evaluation using four metrics: Precision, Recall, F1, and ACC, displayed in Fig.\ref{fig:main-view}(a), \rev{allowing users to quickly comprehend the LLM's general performance at a glance}.

~\\\noindent \textbf{Detail Result View}\hspace{1em}
\label{sec:vis:tab-overall-performance}

This view displays the four metrics in all extracting parameters from synthesis paragraphs by default. As depicted in Fig.\ref{fig:main-view}(b), the heat map uses varied colors to represent the performance of each parameter according to different metrics. Each cell of the heat map represents its corresponding parameter and metric.
Users can alter the visual encoding of the heat map, including rectangular height, circular radius, plain text, and color \& text (used in Fig.\ref{fig:main-view} b), by clicking directly on heat map elements or just through the control panel on the right. This functionality allows for flexible visualization adjustments.

Users can interact with the panel on the right to toggle between metrics and data statistics, showing TP, TN, FP, FN in all parameters.

~\\\noindent \textbf{Extracted Paragraphs View}\hspace{1em}
\label{sec:vis:paragraphs-info}

As depicted in Fig.\ref{fig:main-view}(c), this panel consists of three sections. At the top, a shortcut control bar allows users to sort model-extracted paragraphs either by single selection or direct focus on specific paragraphs using the selection button. Each filtering action triggers an automatic recalculation of the panel data, updating the corresponding visual charts accordingly.

\rev{
The visualization system additionally supports two advanced views: \textbf{Error Map View} and \textbf{Data Distribution View}, that are discretely accessible through tabs in the middle section of the page. This design choice serves multiple purposes: it provides interaction tools of varying complexity for different users while simultaneously addressing potential information overload on the main page.
}

~\\\noindent \textbf{Error Map View}\hspace{1em}

This view (Fig.\ref{fig:main-view} d) provides an intuitive visualization of the relationships between different types of errors. Each node in the graph represents a specific error type, false positive (FP) or false negative (FN), associated with a particular parameter. The colored border represents how many paragraphs have the specific error type and a full circle of border represents the maximum number of errors among all nodes. When a synthesis paragraph contains multiple errors, the graph connects these error nodes pairwise, forming a network that illustrates their interrelations. On each link of the graph, there is a number indicating how many synthesis paragraphs have connected these two nodes due to its errors.
In the upper right corner of the interface, there is a sliding value selection module that is used to block connections that do not exceed the specified value. For example, when the user sets the ``Mask line" value to 3, lines with a line value less than or equal to 3 will not be displayed in the diagram.
When the user hovers over a node, the graph will automatically highlight nodes that directly connect to itself.
This graph also supports drag-and-drop interaction, where users can move the nodes of interest to the center of the canvas to observe the surrounding points more clearly.

~\\\noindent \textbf{Data Distribution View}\hspace{1em}
\label{sec:vis:tab-data-distribution}

As depicted in Fig.\ref{fig:main-view}(e), the panel comprises an interactive scatter plot with an embedded parameter selection panel and a control panel, displaying the two-dimensional distribution of all synthesis paragraphs. This distribution encompasses paragraphs existing in the database and those extracted by LLM. Adjacent to the scatter plot are two pie charts illustrating the parameter distributions of selected paragraphs, with the left chart showing the distribution of LLM-extracted results and the right chart displaying the distribution of other selected points.

We utilize one-hot encoding and PCA dimensionality reduction algorithms to produce the two-dimensional distribution. The parameter selection panel in the center allows users to choose which synthetic parameters should be considered in the embedding calculation. For each selected synthesis parameter in a paragraph, including its names and quantities, we generate a unique vector using one-hot encoding. Each paragraph is then decomposed into its synthetic parameters, and vector representations for each are combined to form a complete paragraph vector. These vectors are subsequently reduced to two dimensions using PCA to illustrate the distribution of paragraphs in the scatter plot.

In the scatter plot, black dots represent existing paragraphs, red dots indicate paragraphs extracted by GPT, and blue dots signify paragraphs both extracted by GPT and existing in the database. The visual design employs larger radii for red and blue dots compared to black dots to highlight new or matched data. A legend in the upper-left corner displays the current number of selected points.

Interactive features of this visual system include tooltips that display detailed paragraph information when a user hovers over a point. Users can zoom in and out using the mouse scroll wheel, with a double-click resetting the zoom level, and can pan across the canvas by clicking and dragging. The control panel allows users to toggle the display of axes and adjust the size of points on the scatter plot.

After clicking the first button in the top-right corner of the scatter plot, the plot enters paragraph selection mode. In this mode, users can draw a rectangular area to select dots for further inspection. When selecting a red point (LLM extraction result) or a blue point (corresponding ground truth), its paired point is automatically selected as well. Users can then choose a parameter of interest through the dropdown list, and the pie charts will update to show the parameter distributions of the LLM-extracted results and other selected paragraphs respectively.

\section{Case Study in Material Science}
\label{sec:case-study}

\rev{In this section, we collaborate with domain experts in MOFs synthesis and use three case studies to evaluate our method, along with 122 well-labeled synthesis paragraphs from our dataset as examples.}

\subsection{Entity Resolution Pipeline}

\begin{figure}[htbp]
    \centering
    \includegraphics[width=0.9\linewidth]{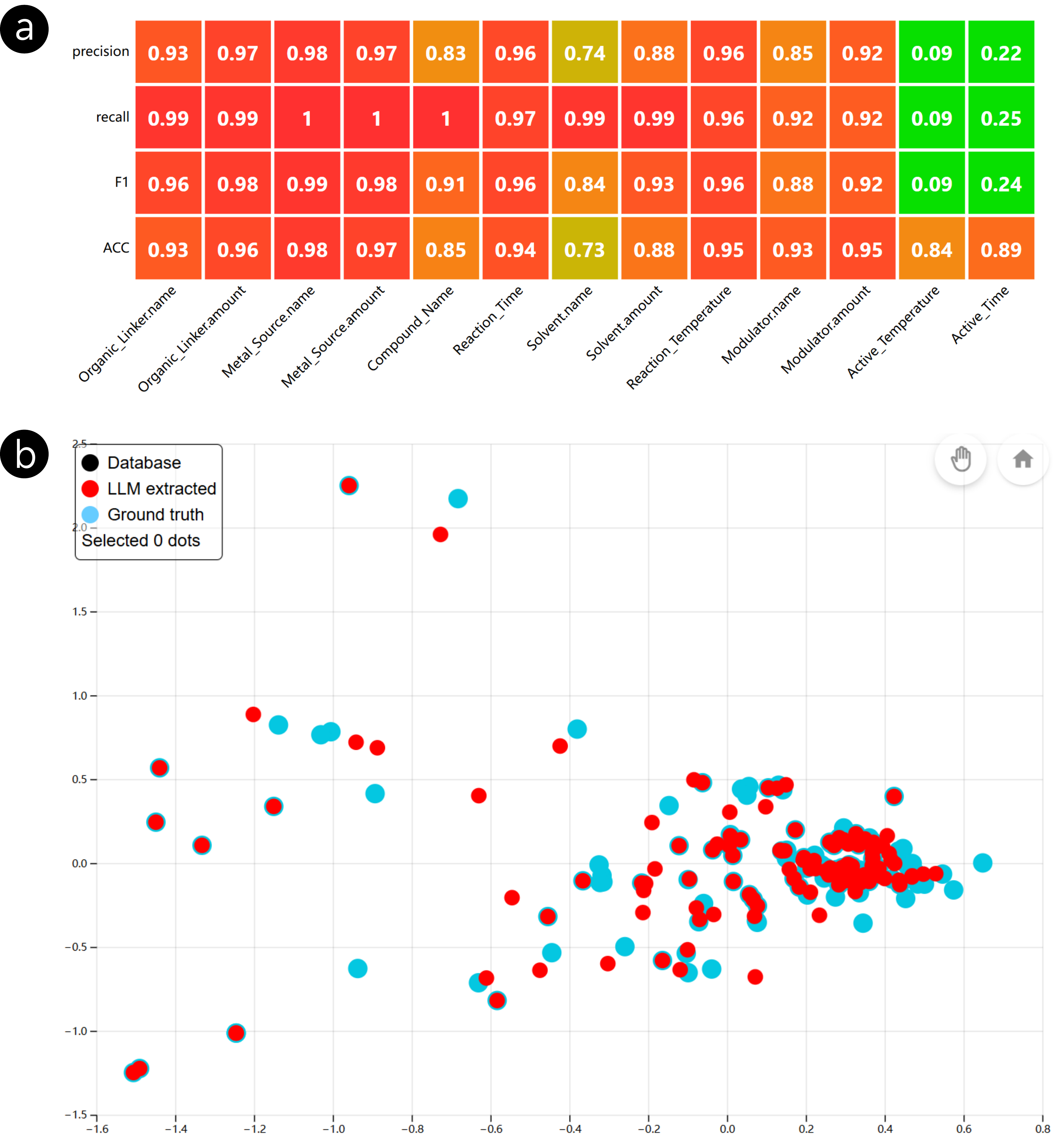}
    
    \caption{Case study of entity resolution pipeline. Chart (a) shows overall metrics of LLM extraction on CSD-MOFs. Chart (b) displays data distribution of all synthesis paragraphs.}
    \label{fig:case-pipeline}
\end{figure}

\rev{Initially, users upload 106 PDF papers, which the framework, equipped with GPT-4, processes through the LLM automatically.
In our test environment, the average time to extract the entities from 122 paragraphs is about 2 minutes.}

Once extraction is complete, users can utilize the filtering panel to select specific paragraphs for analysis. The overall metrics and performance view then displays four key metrics of the LLM resolution, with a default heat map providing a detailed view of entity resolution performance across all evaluation metrics.

The quantitative evaluation result (Fig.\ref{fig:case-pipeline} a) demonstrates the robust performance of our proposed approach \rev{without human intervention}, achieving high scores across multiple metrics: 89.54\% precision, 96.85\% recall, 93.05\% F1-score, and 90.73\% accuracy. The heat map visualization reveals detailed performance variations across different entity types. While the model exhibits strong performance in extracting most parameters, with many achieving above 90\% accuracy, there are notable challenges in certain areas. Specifically, the model shows relatively lower performance in extracting solvent name (0.73 ACC), active temperature (0.84 ACC), and active time (0.89 ACC) parameters, suggesting these entities may require more sophisticated extraction strategies.

To facilitate comprehensive analysis, users can examine the data distribution through an interactive scatter plot visualization (Fig.\ref{fig:case-pipeline} b). The distribution view presents extracted entities as colored points (red and blue), where the proximity between corresponding points indicates the consistency between LLM-extracted results and manual annotations. The high degree of overlap observed in the scatter plot substantiates the reliability of the LLM-based extraction process, with most points showing close spatial correspondence. This visual evidence corroborates the quantitative metrics and provides intuitive validation of the model's extraction capabilities.

Should users decide to replace or re-examine certain paragraphs, they can re-select them in the filtering panel. This action triggers an automatic update of the corresponding performance metrics and visual charts, allowing users to repeat the analysis as needed.

\subsection{LLM Error Analysis}

\begin{figure}[htbp]
    \centering
    \includegraphics[width=0.9\linewidth]{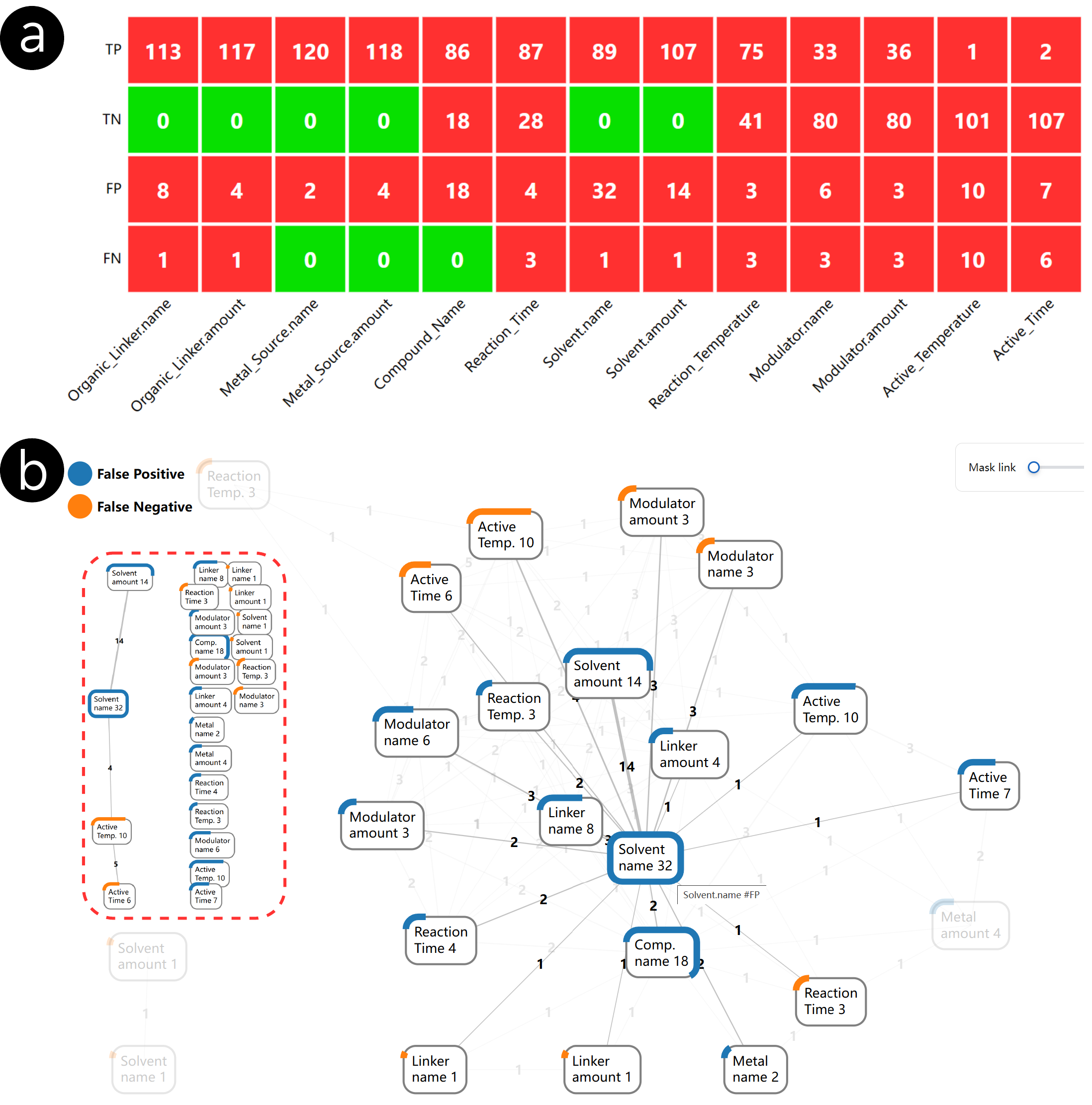}
    
    \caption{Case study of LLM error analysis.
    Chart (a) shows that LLM has poor entity resolution performance for parameters \textit{solvent name}, \textit{solvent amount}, \textit{active temperature} and \textit{active time}.
    Chart (b) shows the error map indicates that there is a strong correlation between \textit{solvent name} and \textit{solvent amount}. When the user set the link mask level to 3 (i.e., links representing no more than 3 synthesis paragraphs will be hidden), it indicates an error pattern of a line from \textit{solvent amount} to \textit{active time}, as shown in the red box.}
    \label{fig:error-anly}
\end{figure}

Detailed error analysis reveals several noteworthy patterns in the LLM's entity extraction performance. Through examination of False Positive (FP) and False Negative (FN) metrics displayed in the count-based heat map (Fig.\ref{fig:error-anly} a), we can identify specific challenges in the model's performance. Most notably, the model exhibits relatively lower accuracy in extracting solvent-related parameters, specifically solvent name and amount.
While other parameters like active time and temperature also show suboptimal performance, this can be attributed to the limited sample size rather than systemic model limitations.

To investigate these solvent-related extraction challenges more thoroughly, we leveraged the error map view for deeper analysis (Fig.\ref{fig:error-anly} b). The visualization revealed a significant correlation in extraction errors, with 14 papers showing concurrent errors in both solvent name and amount parameters.
After setting the linking mask level to 3 (as shown in the red box in the figure), we can observe that errors in \textit{solvent amount}, \textit{solvent name}, \textit{active temperature}, and \textit{active time} formed a line of connection. This strong error correlation pattern suggests a systematic issue in solvent parameter extraction that warrants further investigation.

\begin{figure}[htbp]
    \centering
    \includegraphics[width=0.9\linewidth]{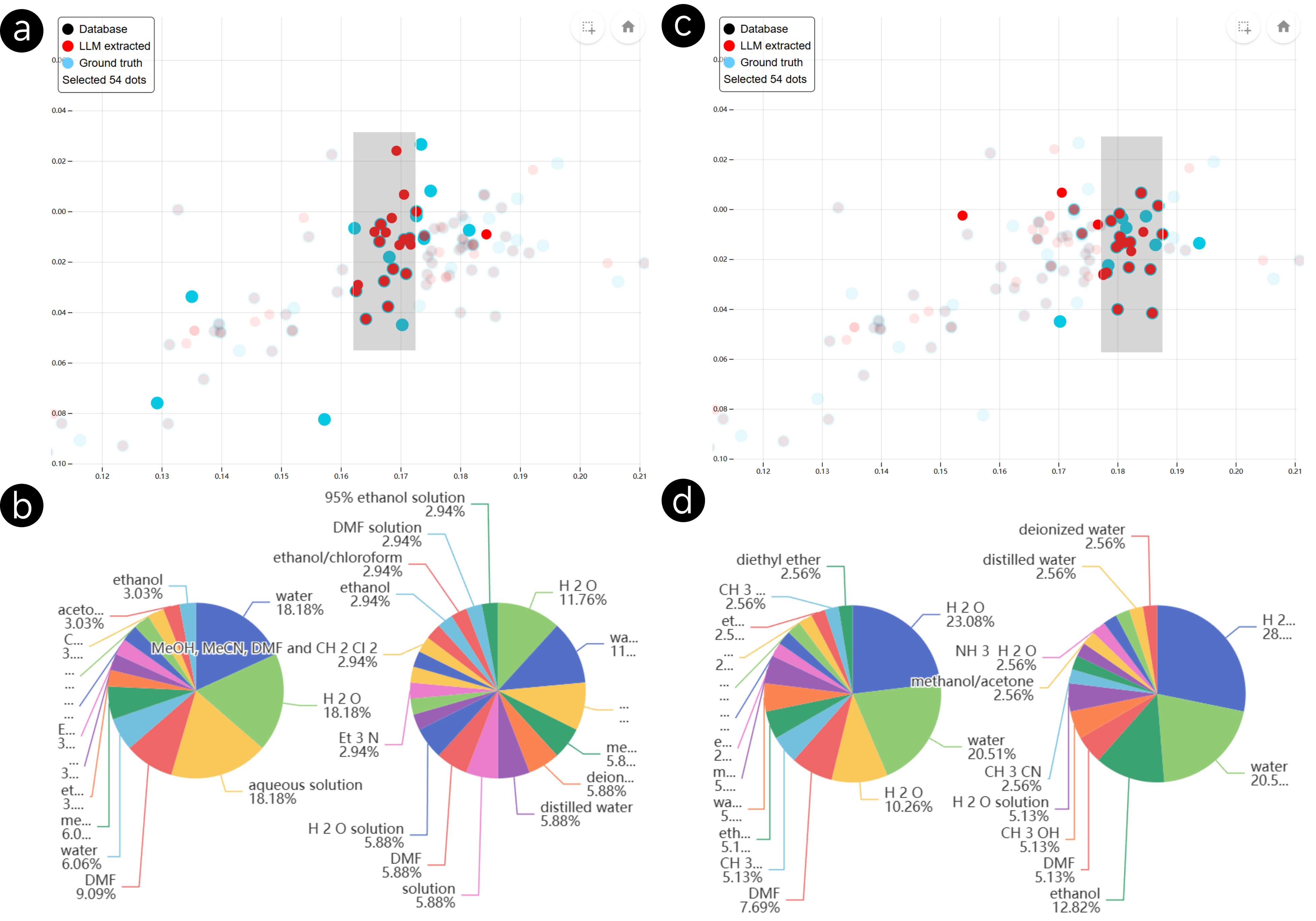}
    
    \caption{Comparison of two groups of synthesis paragraphs on the parameter of solvent name:
    Chart (a) and (b) demonstrate synthesis paragraph extraction results that appear to perform poorly on the scatter plot visually, where the ground truth distribution is observed to be relatively scattered.
    Chart (c) and (d) show another group of data points that appear to perform well, and from the pie chart, their distribution is relatively concentrated.}
    \label{fig:error-anly-d}
\end{figure}

To gain deeper insights into the distribution patterns of these problematic parameters, we utilized the \textbf{Data Distribution View}. By selectively focusing on solvent name and amount parameters through the checkbox filtering mechanism, we generated a focused embedding visualization in the scatter plot. Upon examining the scatter plot, we find a densest cluster of points. We first select 54 points in the left side of the cluster and choose the \textit{solvent name} parameter to draw two pie charts to compare distributions between LLM extracted result and the ground truth, as shown in Fig.\ref{fig:error-anly-d}(a) and Fig.\ref{fig:error-anly-d}(b). We can discover that the dots in the scatter plot shows a relatively bad result and the distribution of \textit{solvent name} is relatively scattered. When we select another 54 points in the right side of the cluster and apply the same analysis, we find that the dots are more compact than the previous group and about 50\% of the solvent is water in this case. We can also find that the two distributions of LLM extracted result are similar.

This multi-faceted error analysis approach, combining count-based heat maps, error correlation mapping, and distribution analysis, provides valuable insights into specific areas where the LLM's extraction capabilities could be improved, particularly in handling common solvent specifications in synthesis protocols.

\subsection{Example Selection in RAG}

\begin{figure}[htbp]
    \centering
    \includegraphics[width=0.9\linewidth]{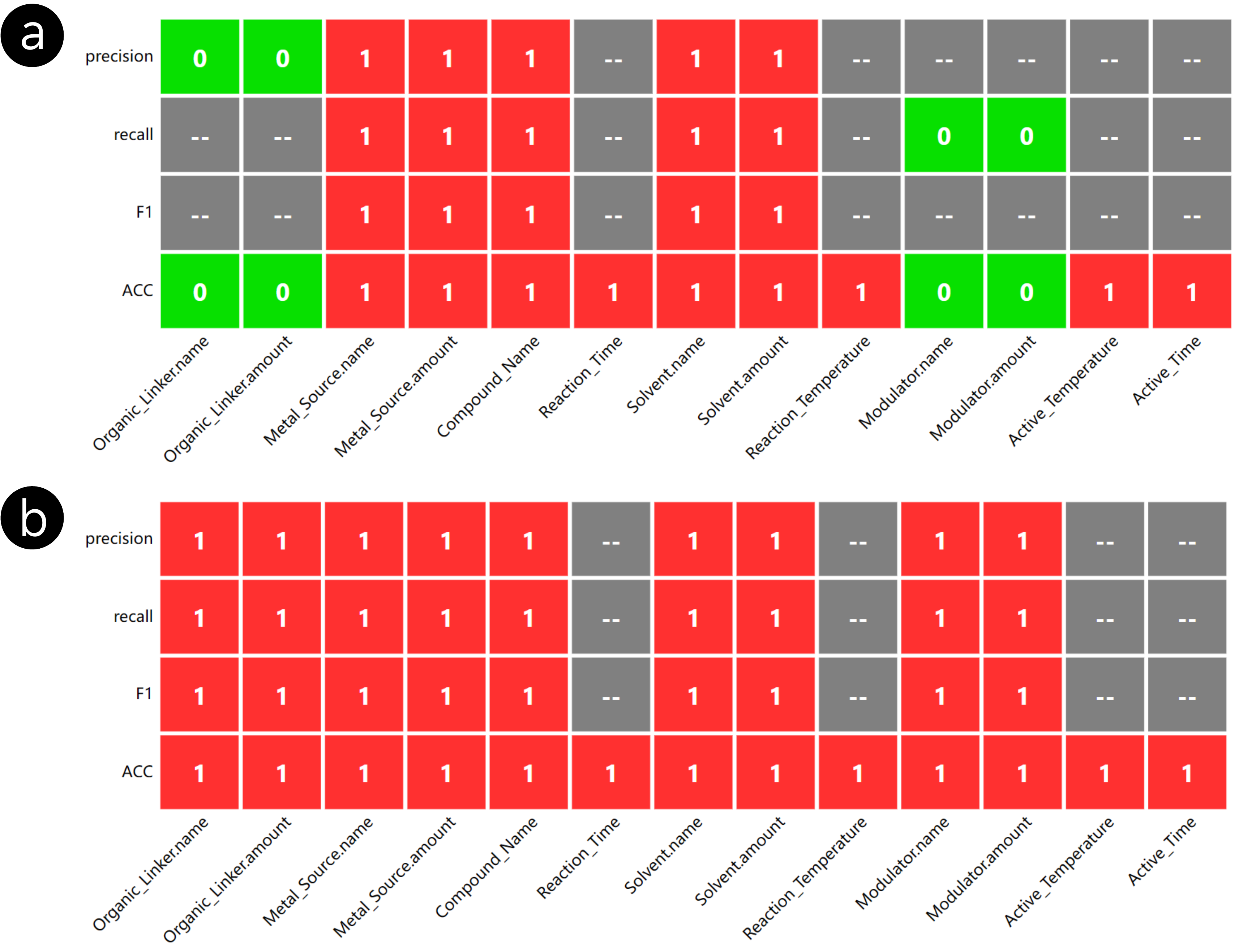}
    
    \caption{Comparison of results using different samples in RAG. Chart (a) is the metric heat map showing the result that LLM produce entity resolution using paragraphs selected by BM25 algorithm automatically, while Chart (b) using manual selected paragraphs for RAG.}
    \label{fig:main-rag-main}
\end{figure}

To demonstrate how different examples of synthesis paragraphs can affect the performance of LLM-based entity resolution, we conducted an experimental case study of one synthesis paragraph. Our analysis reveals the critical role of example selection in RAG (Retrieval-Augmented Generation) for improving entity resolution accuracy.

Initially, when we input the synthesis paragraph into the LLM without careful example selection, the model achieved relatively modest performance metrics: precision, recall, and F1 score were all at 71.43\%, with an overall accuracy of 69.23\% (Fig.\ref{fig:main-rag-main} a). The error map shows that the model struggled particularly with organic linker information and modulator details, indicated by green cells (i.e., the metric value is 0) in these categories.
However, after analyzing the content of the target paragraph and selecting specific synthesis paragraphs through our RAG configuration interface (Fig.\ref{fig:rag}), we observed a dramatic improvement in performance. We identified five key synthesis paragraphs that contained similar structural patterns and experimental conditions (e.g., CdCl$_2$) to our target paragraph. After incorporating these carefully selected examples, the model's performance improved significantly, \rev{with all entities in the ground truth having been extracted}.

The heat map in Fig.\ref{fig:main-rag-main}(b) demonstrates consistent performance across all parameters, with red cells (i.e., the metric value is 1) indicating successful extraction of every entity type. This improvement is particularly notable in previously challenging areas such as organic linker information and modulator details.
This case illustrates that the effectiveness of LLM-based entity resolution is heavily dependent on the quality and relevance of the examples provided through RAG. By selecting synthesis paragraphs that share similar structural patterns and experimental details with the target text, we can significantly enhance the model's ability to accurately identify and extract relevant entities.

\rev{
\section{Discussion and Limitations}
}

\rev{
Through our collaboration with domain experts, we gained several valuable insights about visual analytics for LLM-based entity resolution. Prior to our system, material scientists relied on manual inspection of LLM outputs, resulting in considerable uncertainty about extraction accuracy and limited ability to identify systematic errors. Our visualization design addresses these challenges by opening the ``black box" of LLM extraction in three key ways. First, the hierarchical organization of information (from overall metrics to specific paragraphs) enables users to efficiently navigate between different levels of abstraction without cognitive overload. Second, the interactive error analysis views (e.g., our \textbf{Error Map View} and \textbf{Data Distribution View}) reveal patterns in extraction errors that would be difficult to identify through manual inspection, allowing domain experts to develop targeted strategies for improving extraction performance. Third, the visualization of RAG example selection provides transparency into how different examples influence extraction results, transforming what was previously an opaque process into an interpretable and controllable one.
}

However, there are several design shortcomings in our approach. When users want to compare different extraction parameters, they need to return to the extraction task selection list and re-select the extraction task to view the performance charts. This process becomes cumbersome when conducting iterative experiments or performing comparative analysis.
Moreover, our current RAG implementation lacks comprehensive visualization capabilities to understand the impact of example selection. The system does not support simultaneous comparison of different RAG configurations, making it difficult to identify optimal example combinations. Although we can observe overall performance improvements, users cannot easily track how individual examples contribute to the extraction results or understand the relative importance of different synthesis paragraphs in the RAG process. \rev{Also, scalability is a concern because we only test it on a relatively small dataset of 122 examples.}

Future work could put efforts on improving user interaction, and a fully labeled dataset is required to drive LLMs for better entity resolution.

\section{Conclusion}
\label{sec:conclusion}

This paper demonstrates an effective integration of large language models (LLMs) into a comprehensive visual entity resolution framework for scientific literature analysis. We have developed a fully interactive framework that enables batch entity resolution without requiring programming expertise, significantly improving data interpretation through advanced visualization techniques. Our method utilizes prompt engineering and structured outputs to improve the accuracy and reliability of LLMs, minimizing typical errors in automated entity resolution. The visualization design of the system offers intuitive and detailed data representations, facilitating deep insights and allowing users to interactively adjust model parameters to suit specific research requirements. This study underscores the transformative potential of LLMs in scientific research, paving the way for future developments in automated literature analysis and entity resolution, \rev{and it is easy to transfer this pipeline to similar entity resolution tasks in other domains}.

\section{CRediT Authorship Contribution Statement}
\label{sec:credit}

\textbf{Siyu Wu:} Writing - original draft, Writing - review \& editing, Software, Visualization.

\textbf{Yi Yang:} Writing - original draft, Project administration, Investigation.

\textbf{Weize Wu:} Software.

\textbf{Ruiming Li:} Software.

\textbf{Yuyang Zhang:} Software.

\textbf{Ge Wang:} Conceptualization.

\textbf{Huobin Tan:} Conceptualization.

\textbf{Zipeng Liu:} Writing - review \& editing, Conceptualization.

\textbf{Lei Shi:} Writing - review \& editing, Conceptualization.

\section{Declaration of Competing Interest}
\label{sec:comp-interest}

All authors declare that they have no competing personal relationships or financial interests that could have appeared to influence the work reported in this paper.

 \bibliographystyle{elsarticle-harv} 
 \bibliography{reference}

\end{document}